\documentclass[twocolumn,english,prd]{revtex4}
\usepackage{graphicx}
\begin{document} 
\newcommand{\Tr}{\mbox{Tr\,}}
\newcommand{\beq}{\begin{equation}}
\newcommand{\eeq}{\end{equation}}
\newcommand{\bea}{\begin{eqnarray}}
\newcommand{\eea}{\end{eqnarray}}
\newcommand{\bt}{\begin{tabular}}
\newcommand{\et}{\end{tabular}}
\newcommand{\kvec}{{\bf k}}
\newcommand{\ivec}{{\bf i}}
\newcommand{\jvec}{{\bf j}}
\newcommand{\smax}{\mbox{\tiny{max}}}
\newcommand{\q}{{\bf q}}
\newcommand{\p}{{\bf p}}
\newcommand{\de}{{\rm d}}
\newcommand{\mnras}{Mon. Not. R. Astron. Soc.}
\newcommand{\aap}{Astron. Astrophys.}
\newcommand{\apjs}{Astrophys.\ J.\ Suppl.\  }
\newcommand{\apjl}{Astrophys.\ J.\ Lett.\  }
\newcommand{\physrep}{Phys.\ Rep.\  }
\def\d{\delta}
\def\Mpc{\, h^{-1} \, {\rm Mpc}}
\def\Gpc{\, h^{-1} \, {\rm Gpc}}
\def\Gpccube{\, h^{-3} \, {\rm Gpc}^3}
\def\kvecMpc{\, h \, {\rm Mpc}^{-1}}
\def\la{\mathrel{\mathpalette\fun <}}
\def\ga{\mathrel{\mathpalette\fun >}}
\def\fun#1#2{\lower3.6pt\vbox{\baselineskip0pt\lineskip.9pt
        \ialign{$\mathsurround=0pt#1\hfill##\hfil$\crcr#2\crcr\sim\crcr}}}

\title{Galaxy Bias and Halo-Occupation Numbers from Large-Scale Clustering}
\author{Emiliano Sefusatti$^1$ and  Rom\'an Scoccimarro$^{1,2}$}
\vskip 2pc
\address{${}^{1}$Center for Cosmology and Particle Physics, \\
Department of Physics, New York University
New York, NY 10003 \\
${}^{2}$Kavli Institute for Cosmological Physics, University of Chicago,
Chicago, IL 60637
}

\begin{abstract}

We show that current surveys have at least as much signal to noise in higher-order statistics as in the power spectrum at weakly nonlinear scales. We discuss how one can use this information to determine the mean of the galaxy halo occupation distribution (HOD) using only large-scale information, through galaxy bias parameters determined from the galaxy bispectrum and trispectrum. After introducing an averaged, reasonably fast to evaluate, trispectrum estimator, we show that the expected errors on linear and quadratic bias parameters can be reduced by at least $20$-$40$\%. Also, the inclusion of the trispectrum information, which is sensitive to "three-dimensionality" of structures, helps significantly in constraining the mass dependence of the HOD mean. Our approach depends only on adequate modeling of the abundance and large-scale clustering of halos and thus is independent of details of how galaxies are distributed within halos. This provides a consistency check on the traditional approach of using two-point statistics down to small scales, which necessarily makes more assumptions. We present a detailed forecast of how well our approach can be carried out in the case of the SDSS.

\end{abstract}

\maketitle

\section{Introduction}

Galaxy clustering in current surveys is providing tight constraints on cosmological parameters and the nature of primordial fluctuations~\cite{Percival2004,Tegmark2004}. One of the major issues in obtaining  constraints from galaxy surveys on cosmology is the relationship between galaxy clustering and the underlying dark matter. This ``galaxy bias" can be best studied at large scales using higher-order statistics~\cite{Frieman:1993nc,Fry:1992vr}, such as the higher-order moments~\cite{Szapudi02,Croton04}, the three-point function~\cite{Frieman:1999qj,Gazta02,JiBo04, Kayo04}, and the bispectrum~\cite{Fry1994,Scoccimarro:2000sp,Feldman:2000vk,Verde:2002ed}. 

In this paper we consider how well one can measure the galaxy trispectrum, and how this can be used together with the bispectrum to place constraints on galaxy bias at large scales. The trispectrum is the lowest order statistic that is sensitive to the three-dimensional character of structures generated by gravitational instability and thus is a natural candidate to tell us interesting new information not contained in the power spectrum and bispectrum. Here we show that in surveys currently under completion, there will be enough signal to noise in the galaxy trispectrum to provide improved constraints over measurements of the bispectrum alone on galaxy linear and nonlinear bias parameters.  Four-point statistics have so far only been measured mainly at small scales in angular catalogs~\cite{FrPe78,MSS92,SSB92,SDES95}, with a marginal detection of the trispectrum in redshift surveys~\cite{BaFr91}. The use of the {\em disconnected} (Gaussian) part of the trispectrum (not the one that concerns us here) to probe primordial non-Gaussianity is studied in~\cite{VeHe01}. Previous estimates of the accuracy of higher-order moments and three-point statistics expected in current surveys are given in~\cite{MVH97,CSS98,SCB99,SSZ04}.

\begin{figure*}
\begin{center}
\begin{tabular}{cc}
{\includegraphics[width=0.5\textwidth]{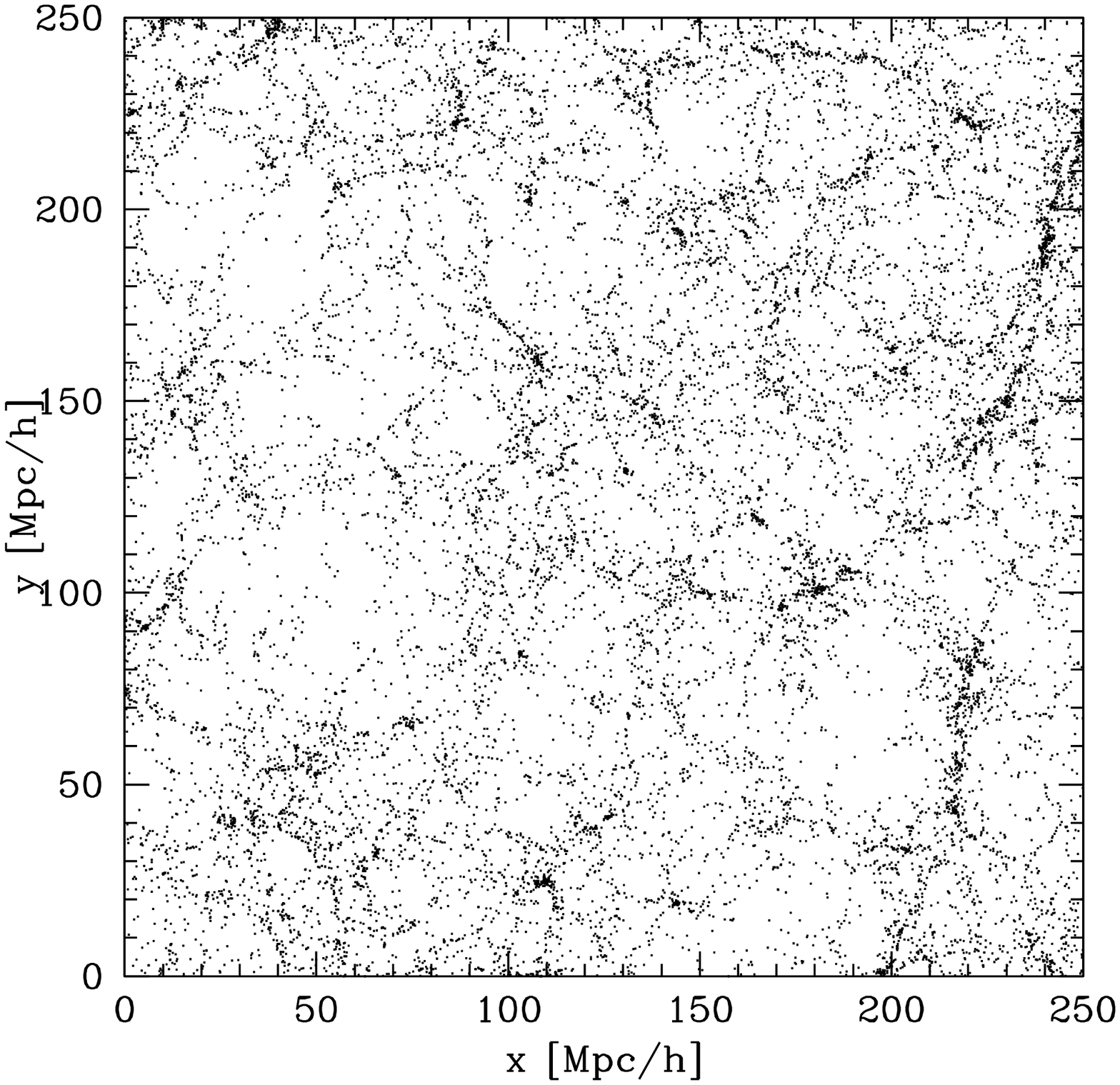}}&
{\includegraphics[width=0.5\textwidth]{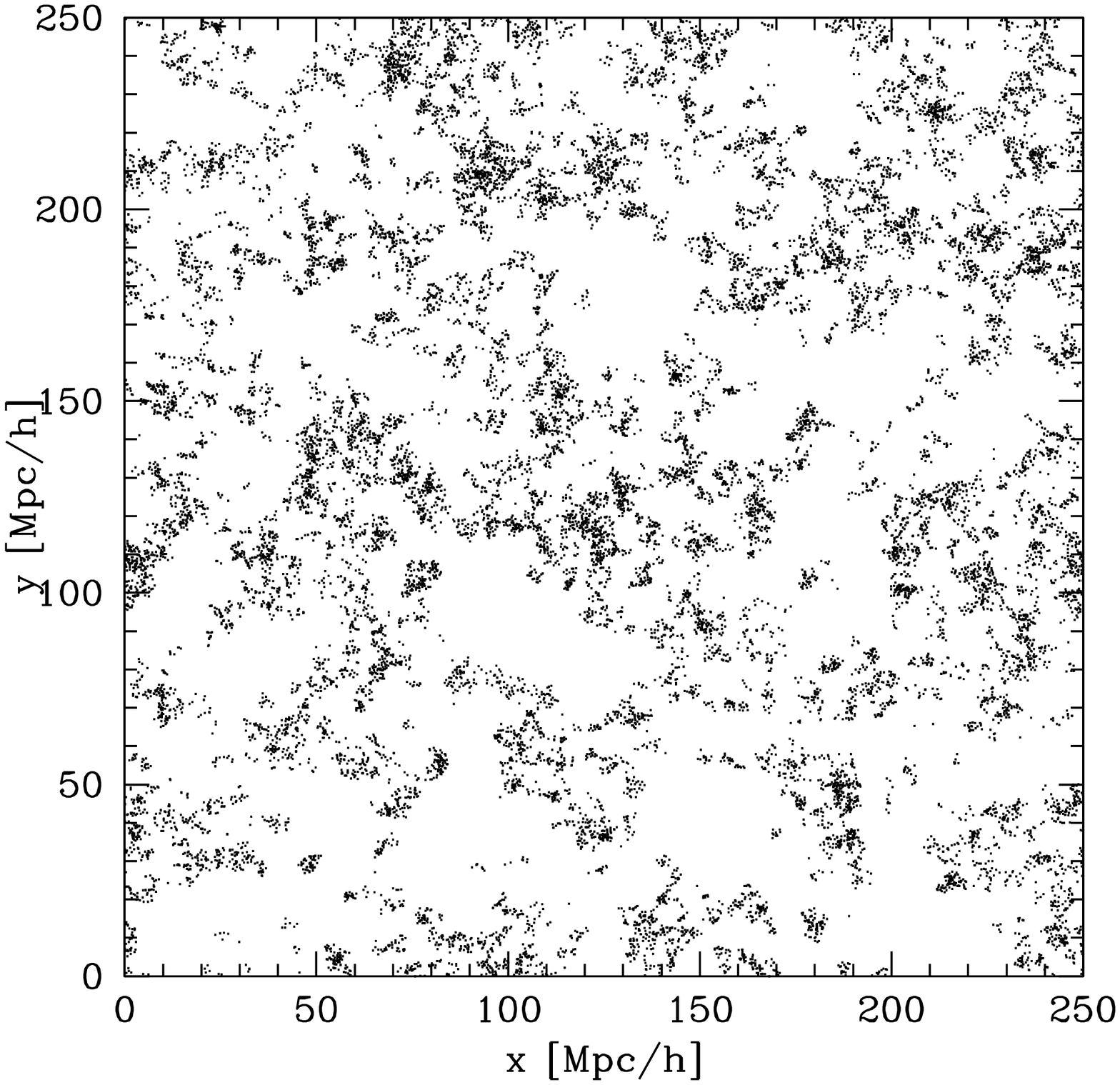}}
\end{tabular}
\caption{Slices 50 Mpc/$h$ thick of a mock galaxy distribution obtained from an HOD fit in a $\Lambda$CDM model to the $M_r<-20$ galaxy two-point function in SDSS  (left) and a Rayleigh-L\`evy flight (right). Despite their obvious differences, these two distributions have the same two-point statistics, the differences seen are entirely due to those in higher-order correlations, see Fig.~\protect\ref{PkQBQT}.}
\label{HODRLwalkdist}
\end{center}
\end{figure*}
\begin{figure*}
\begin{center}
\begin{tabular}{cc}
{\includegraphics[width=0.5\textwidth]{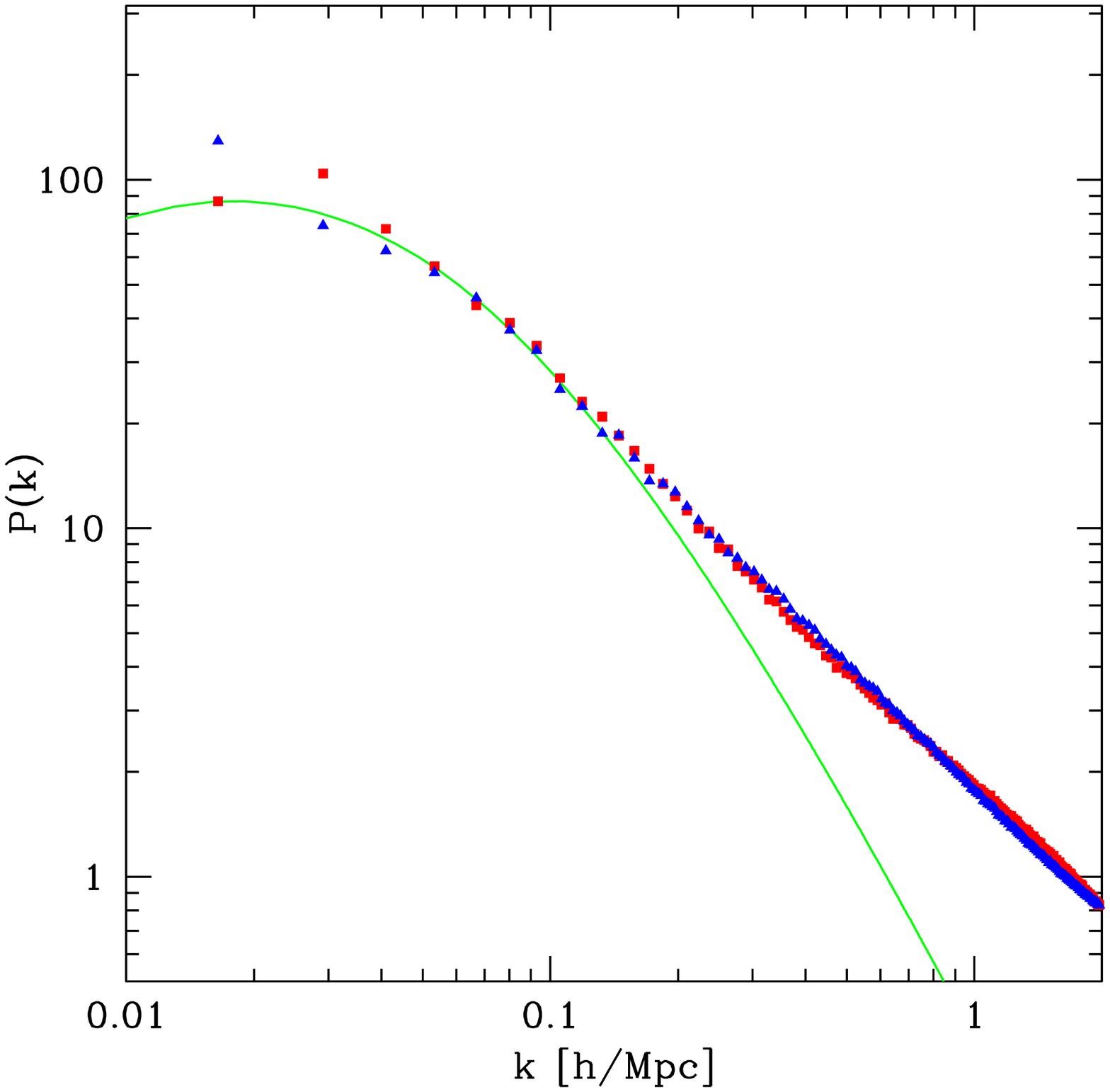}}&
{\includegraphics[width=0.5\textwidth]{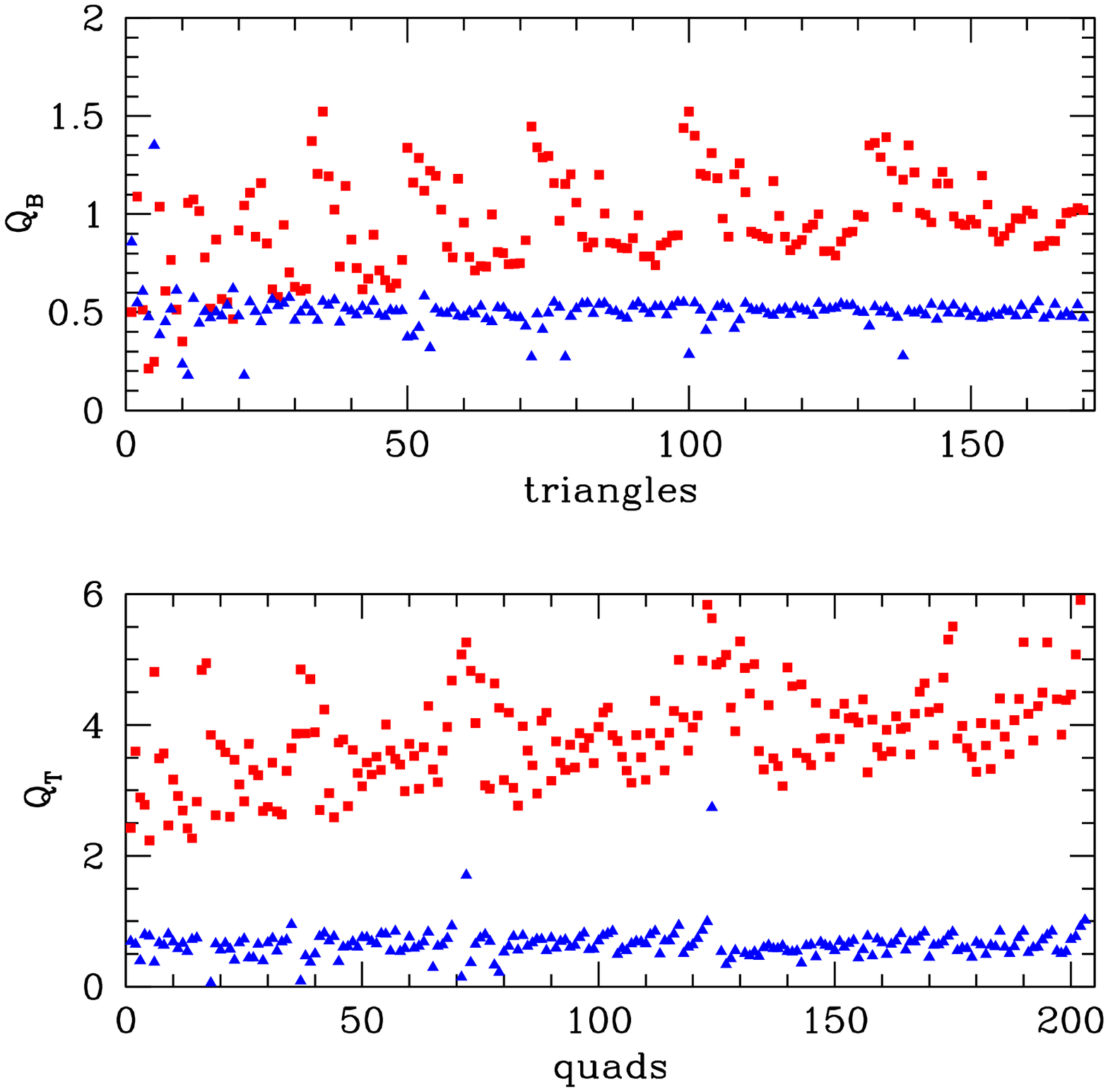}}
\end{tabular}
\caption{The distributions in Fig.~\protect\ref{HODRLwalkdist} have the same power spectrum (left) but can be easily distinguished by their bispectrum  and trispectrum (right). Square symbols correspond to the HOD galaxies, triangles to the Rayleigh-L\`evy flight. The bispectrum ($Q_B$) and trispectrum ($Q_T$) are for all shapes of triangles and ``quads" (see section~\protect\ref{BTestim}) in the range $0.04 \kvecMpc \leq k \leq 0.4 \kvecMpc$, here binned into $N_T=170$ and $N_Q=203$ configurations, respectively. The variations seen in $Q_B$ and $Q_T$ in the HOD galaxies are due to the dependence of higher-order correlations on the shape of the configuration, a reflection of the filamentary structure seen in the left panel in Fig.~\protect\ref{HODRLwalkdist}.}
\label{PkQBQT}
\end{center}
\end{figure*}

Galaxy biasing is at present best connected to galaxy formation using the ``halo model", where galaxies are only present within dark matter halos in numbers prescribed by a halo occupation distribution (HOD) and with a profile dictated by numerical simulations~\cite{Ma:2000ik,Seljak:2000gq,Scoccimarro:2000gm, Benson01,White01,BeWe02, Zheng02,Berlindetal,Cooray:2002di,TaJa03,Kravtsov:2003sg,Zentner:2003yd}. In this language, it is possible to directly map the large-scale bias parameters into a probe of the mean of the HOD using only information about the mass function and large-scale bias of dark matter halos~\cite{Scoccimarro:2000gm}, which depends on better understood large-scale physics. In this paper we address how well one can constrain the mean of galaxy halo-occupation numbers from measurement of the bias parameters at large scales. 

Our approach to constrain the mean of the HOD is complementary to constraints based on measurements of two-point statistics down to small scales~\cite{Scranton03,Zehavi:2004ii,Zheng:2004id,vandenBosch:2004mh,Kev04} where more details need to be modeled, such as  halo exclusion, galaxy distribution profiles inside halos, velocity bias between galaxies and dark matter, and the second moment of the HOD distribution. Therefore, our method can provide a consistency check on these additional assumptions needed for small-scale studies. In addition, because higher-oder statistics measure the large scale linear bias this breaks degeneracies between bias, $\Omega_m$ and $\sigma_8$ which is necessary in order to predict the halo mass function and halo bias that maps constraints on bias into constraints on the mean of the HOD.  Therefore, the halo statistics and the mean HOD are determined simultaneously. Other work has also obtained constraints on HOD parameters from higher-order statistics going down to small scales~\cite{Scoccimarro:2000gm, PTHalos,Wangetal04}, but our purpose here is to see how much can be done with large-scale information where only the simpler physics of halos plays a role.

From a physical point of view, measuring three and four-point statistics at large scales gives a rather complete picture of the non-linear couplings induced during the formation of large-scale structures, and thus a fundamental test of gravitational instability~\cite{PTreview}. From a purely statistical point of view, the richness in the dependence of the bispectrum and trispectrum on configuration of the points allows to disentangle the relative probabilities of elongated versus compact shapes (bispectrum) and planar versus three-dimensional character of large-scale structures (trispectrum). 

That higher-order statistics can help to break degeneracies otherwise present should be of no surprise,  although a visual example may illustrate the power in this method more clearly. Figure~\ref{HODRLwalkdist} shows two distributions that clearly look ``very different" to the eye. The left panel shows a mock galaxy distribution obtained from an HOD fit to the $M<-20$ galaxy two-point function in SDSS~\cite{Zehavi:2004ii} assuming a $\Lambda$CDM halo population. The right panel shows instead a Rayleigh-L\`evy flight~\cite{1975CRASM.280.1551M,1980lssu.book.....P,SC96} with parameters chosen to match the power spectrum of the previous distribution~\footnote{In this model there are three parameters, the number of ``clusters", how many objects (random walks) constitute a cluster, and a spectral index characterizing the power-law decay of the distance of each step of the walk, see~\protect\cite{1980lssu.book.....P}.}. 

The left panel in Fig.~\ref{PkQBQT} shows that indeed the power spectra {\em at all scales} are very similar, and thus degenerate. That this can happen should not be too surprising, after all the two-point function (or power spectrum) only measures the average number of neighbors from a given object as a function of separation, a rather crude statistic.  The right panel in Fig.~\ref{PkQBQT} shows that the two distributions are easily distinguished by their bispectrum (top) and trispectrum (bottom) for essentially all configurations of points. The Rayleigh-L\`evy flight predicts $Q_B=0.5$ and $Q_T\simeq0.75$ independent (approximately for $Q_T$) of configuration and scale~\cite{1980lssu.book.....P,SC96}~\footnote{Note that our definition of $Q_T$ in Eq.~(\protect\ref{QDMdef}) is not standard, since we don't include all possible combinations in the denominator. Doing that leads to $Q_N=2^{1-N}N!/N^{N-2}$ for the $N-$point case~\protect\cite{SC96}.}. It is interesting to note that this model was proposed in the 70's in response  to the observational results from the Lick catalog that showed  $Q_B,Q_T$ being consistent with constants at small scales; this {\em ruled out} the previous incarnation of the halo model~\cite{Peebles74,MCSilk77}, where galaxies populate identical halos with power-law profiles chosen to match two-point statistics.

This paper is organized as follows. In the next section we briefly review the bispectrum and trispectrum generated by gravitational instability at large scales, the effects on it of galaxy biasing and the estimators of the bispectrum and trispectrum. In section~\ref{DetB} we discuss the determination of bias parameters from galaxy surveys and compare the signal to noise in higher-order statistics to that in the power spectrum. Finally, in section~\ref{biTOhod} we show how one can turn the constraints on bias parameters into constraints on the mean HOD.

\section{The LSS  Bispectrum and Trispectrum}
\label{LSSBT}

\subsection{Bispectrum and Trispectrum generated by Gravity at Large Scales}

In this paper we will assume the dark matter primordial fluctuations to be Gaussian. The three-point function and the connected four-point function observed in galaxy surveys will then be a consequence of gravitational instability and galaxy biasing. At the scales relevant for this study, we can work in Eulerian Perturbation Theory (EPT) taking into account corrections to linear evolution $\delta^L$ of second-order for the bispectrum and up to third-order corrections for the trispectrum, 
\begin{eqnarray}
\langle\delta_{\kvec_1}\delta_{\kvec_2}\rangle\!\!\! & \equiv & \!\!\!\delta_D (\kvec_{12})\  P(k_1),\\
\langle\delta_{\kvec_1}\delta_{\kvec_2}\delta_{\kvec_3}\rangle\!\!\! & \equiv &\!\!\! \delta_D (\kvec_{123})\ B(k_1,k_2,k_3),\\
\langle\delta_{\kvec_1}\delta_{\kvec_2}\delta_{\kvec_3}\delta_{\kvec_4}\rangle_c\! \!\!& \equiv &\!\!\! \delta_D (\kvec_{1234}) T(\kvec_1,\kvec_2,\kvec_3,\kvec_4),
\end{eqnarray}
where $\langle ...\rangle_c$ implies that only \emph{connected} terms are included in the average,
\beq
B(k_1,k_2,k_3) = 2 F_2(\kvec_1,\kvec_2)P_1 P_2+\;\text{cyc.},
\eeq
is the bispectrum, and the trispectrum can be split into two different contributions, $T=T_a+T_b$ with
\bea\label{TaTb}
T_a&=&\!4 P_1 P_2 \left[ P_{13} F_2(\kvec_1,-\kvec_{13}) F_2(\kvec_2,\kvec_{13})\right. \nonumber \\
&+ &\left.  P_{14} F_2(\kvec_1,-\kvec_{14}) F_2(\kvec_2,\kvec_{14})\right]+\,{\rm cyc.},\\
T_b &=& \! \left[F_3(\kvec_1,\kvec_2,\kvec_3)+\!{\rm perm.}\right] P_1 P_2 P_3 +\!\rm{cyc.},
\eea
where $P_i = P(k_i), P_{ij}=P(|\kvec_i+\kvec_j|)$. It is useful to introduce the reduced bispectrum $Q_B$ and trispectrum $Q_T$,
\beq\label{defQB} 
Q_B(k_1,k_2,k_3)\equiv\frac{B(k_1,k_2,k_3)}{P_1 P_2+P_1 P_3+P_2 P_3},
\eeq
\beq\label{QDMdef}
Q_T(\kvec_1,\kvec_2,\kvec_3,\kvec_4)\equiv\frac{T(\kvec_1,\kvec_2,\kvec_3,\kvec_4)}{P_1 P_2 P_3 +\; {\rm cyc.}}
\eeq
which have the advantage of being almost independent of scale and cosmological parameters such as $\Omega_m$ and $\sigma_8$. 

The $F_2$ and $F_3$ kernels describe the second and third order solutions in EPT, and can be written in terms of two fundamental mode-coupling functions,
\beq
\alpha(\kvec_1,\kvec_2)=\frac{\kvec_{12}\cdot\kvec_1}{k_1^2},
\eeq
\beq
\beta(\kvec_1,\kvec_2)=\frac{k_{12}^2(\kvec_1\cdot\kvec_2)}{2 k_1^2 k_2^2},
\eeq
which represent the nonlinearities involved in mass and momentum conservation, respectively. The relationship between them and the kernels read,
\bea
F_2 &=& \frac{5}{7}\alpha(\kvec_1,\kvec_2)+\frac{2}{7} \beta(\kvec_1,\kvec_2)
\nonumber \\ & = &\frac{5}{7}+\frac{x}{2}\left(\frac{k_1}{k_2}+\frac{k_2}{k_1}\right)
+\frac{2}{7}\ x^2,
\label{F2}
\eea

where  ($x \equiv \hat{k}_1 \cdot \hat{k}_2$) and 
\bea
F_3\!&\! =\! &\! \frac{7}{18}\ \alpha(\kvec_1,\kvec_{23})\ F_2(\kvec_2,\kvec_3)
\nonumber \\ 
\!&\! +\! &\! 
\frac{1}{9}\ \beta(\kvec_1,\kvec_{23})\ G_2(\kvec_2,\kvec_3) 
\nonumber \\
\!&\!+\!&\! 
\frac{1}{18}G_2(\kvec_1,\kvec_2)\left[7\alpha(\kvec_{12},\kvec_3)+
2\beta(\kvec_{12},\kvec_3)\right], 
\nonumber \\  & &
\eea
where the kernel $G_2$ is obtained from $F_2$ in Eq.~(\ref{F2}) by replacing 5 by 3 and 2 by 4. We thus see that in a sense, the bispectrum and trispectrum have a rather ``complete" information of large-scale clustering, in principle one could try to deduce $\alpha$ and $\beta$ from $B$ and $T$. We shall explore this possibility in future work~\cite{trispectrum}.

\subsection{Galaxy Biasing at Large Scales}

Since gravity is the only long-range force in the problem, at large scales we can assume the bias to be local, therefore when smoothed over large enough scales (lcompared to dark matter halo sizes) the galaxy number density contrast and  dark matter density contrast are related by \cite{Fry:1992vr}
\beq\label{bias}
\delta_g \simeq b_1\delta+\frac{b_2}{2}\delta^2+\frac{b_3}{6}\delta^3
\eeq
with $b_1$, $b_2$ and $b_3$ are constants, the \emph{bias parameters}. The galaxy power spectrum at large scales is then given by $P^{(g)}(k)\simeq b_1^2 P(k)$, while the bispectrum of the galaxy distribution can be expressed in terms of the dark matter bispectrum as 
\bea\label{bias_bisp}
B^{(g)}(k_1,k_2,k_3) & = & b_1^3\ B(k_1,k_2,k_3)
\nonumber\\
& + & b_1^2 b_2\ (P_1 P_2+ \textrm{cyc.})
\eea
while the reduced galaxy bispectrum $Q_B^{(g)}$ is
\beq
Q_B^{(g)}=\frac{Q_B}{b_1}+\frac{b_2}{b_1^2}.
\label{Qgtot}
\eeq
For the galaxy trispectrum it follows that 
\bea\label{Tg}
T^{(g)} &=& b_1^4\,T^{(1)}+\frac{b_1^3b_2}{2}\,T^{(2)}
\nonumber\\
&+& \frac{b_1^2b_2^2}{4}\,T^{(3)}+\frac{b_1^3b_3}{6}\,T^{(4)}
\eea
where $T^{(1)}=T_a+T_b$ as defined in Eq.~(\ref{TaTb}) and, 
\bea
T^{(2)} & = & 4 P_1 \left[ F_2(\kvec_2,\kvec_3)P_2 P_3 + F_2(\kvec_2,-\kvec_{23}) P_2 P_{23}\right.  \nonumber\\
& +& \left. F_2(\kvec_3,-\kvec_{23})P_3 P_{23} \right]+ 4 P_2 \left[F_2(\kvec_1,\kvec_3)P_1 P_3\right. 
 \nonumber\\
& +&\left. F_2(\kvec_1,-\kvec_{13}) P_1 P_{13}+F_2(\kvec_3,-\kvec_{13})P_3 P_{13}\right]\nonumber\\
& +&  4 P_3 \left[F_2(\kvec_1,\kvec_2)P_1 P_2 + F_2(\kvec_1,-\kvec_{12}) P_1 P_{12}\right. \nonumber\\
& + & \left.F_2(\kvec_2,-\kvec_{12})P_2 P_{12} \right]+\,\rm{cyc.} \quad[\rm{4\;terms}]\\
T^{(3)} & = & 4 P_1 P_2 \left(P_{13} + P_{14} \right)+\,\rm{cyc.} \quad[\rm{12\;terms}]\\
T^{(4)} & = & 6 P_1 P_2 P_3 +\,\rm{cyc.} \quad[\rm{4\;terms}]
\eea
The \emph{reduced} trispectrum, in this case, reads
\beq\label{Qttot}
Q_T^{(g)}=\frac{1}{b_1^2}\,Q^{(1)}_T+\frac{b_2}{2 b_1^3}\,Q^{(2)}_T+\frac{b_2^2}{4 b_1^4}\,Q^{(3)}_T+\frac{b_3}{b_1^3}
\eeq
where $Q^{(i)}_T=T^{(i)}/(P_1 P_2 P_3 +\,\rm{cyc.})$ and $Q^{(4)}_T = 6$ by definition.

\subsection{Bispectrum and Trispectrum estimators}
\label{BTestim}
 
To discuss our particular implementation of an averaged trispectrum estimator we note that given Fourier coefficients a bispectrum estimator can be written as~\cite{Scoccimarro:1997st}

\beq
\hat{B}_{123}\equiv  \frac{V_f}{V_B}\int_{k_1}\!\!\!\!d^3 q_1\!\!\int_{k_2}\!\!\!\!d^3 q_2\!\!\int_{k_3}\!\!\!\!d^3 q_3 \delta_D(\q_{123})\delta_{\q_1}\delta_{\q_2}\delta_{\q_3}, \label{Best}
\eeq
where the integration is over the bin defined by $q_i\in(k_i-\delta k/2,k_i+\delta k/2)$, $V_f\equiv k_f^3=(2\pi)^3/V$ is the volume of the fundamental cell in Fourier space, and
\bea
V_B &\equiv& \int_{k_1} \!\!\!\! d^3 q_1\int_{k_2} \!\!\!\! d^3 q_2 \int_{k_3} \!\!\!\! d^3 q_3 \,\delta_D(\q_{123})
\nonumber\\
 &\simeq& 8\pi^2\ k_1 k_2 k_3\ \delta k^3. \label{VB}
\eea
The corresponding variance is 
\beq
\Delta \hat{B}^2_{123}=V_f \frac{s_B}{V_B}\ P_{tot}(k_1)P_{tot}(k_2)P_{tot}(k_3)
\label{Berror}
\eeq
where $s_B=6,2,1$ for equilateral, isosceles and general triangles, respectively, and the total power spectrum (accounting for shot noise),
\beq
P_{tot}(k)  \equiv P(k)+\frac{1}{(2 \pi)^3}\ \frac{1}{ \bar{n}}.
\eeq
Such a definition for an estimator is trivially extended to the trispectrum. A particular configuration of the 4-point function is completely determined given $6$ parameters ($N(N-1)/2$ for the $N$-point case). These can be, for instance, the four lengths $k_i\equiv |\kvec_i|$ for $i=1,2,3,4$ plus the diagonals $d_1\equiv|\kvec_1-\kvec_2|$ and $d_2\equiv|\kvec_1-\kvec_3|$. The trispectrum estimator can then be written as
\bea
\hat{T} & \equiv & \frac{V_f}{V_T}\int_{k_1}\!\!\!\!d^3 q_1\ldots \int_{k_4}\!\!\!\!d^3 q_4\int_{d_1}\!\!\!\!d^3 p_1\int_{d_2}\!\!\!\!d^3 p_2 \;\delta_D(\q_{1234})\nonumber\\
& & \times\,\delta_D(\p_1-\q_1+\q_2) \,\delta_D(\p_2-\q_1+\q_3) \nonumber \\ & & \times \,\delta_{\q_1}\delta_{\q_2}\delta_{\q_3}\delta_{\q_4} \label{Test}
\eea
where the integrations are taken over bins which are spherical shells in Fourier space of thickness $\delta k$, and $V_T$ denotes the same integral as in the numerator but with Fourier coefficients replaced by one, as in the bispectrum case [see Eqs.~(\ref{Best}-\ref{VB})].
However, in this work we will pursue a simpler approach, leaving the more detailed general case for a future paper. One can construct an ``angle averaged'' trispectrum that depends only on 4 variables, rather than 6 by removing the constraint on the diagonals, i.e. 
\bea
\widetilde{T}_{1234} & \equiv & \frac{V_f}{V_{\widetilde{T}}}
\int_{k_1}\!\!\!\!d^3 q_1\cdots\int_{k_4}\!\!\!\!d^3 q_4\;\delta_D(\q_{1234}) 
 \delta_{\q_1}\delta_{\q_2}\delta_{\q_3}\delta_{\q_4}, \nonumber \\ & & 
\label{Testim}
\eea
where $V_{\widetilde{T}}$ is given by
\bea
V_{\widetilde{T}} & \equiv & \int_{k_1}\!\!\!\!d^3 q_1\cdots\int_{k_4}\!\!\!\!d^3q_4\;\delta_D(\q_{1234})  \nonumber \\& = & 8\pi^3\delta k^4\  k_1\, k_2 \, k_3 \, k_4\nonumber \\ & & \times
 \Big( k_1+k_2+k_3+k_4 -|k_1+k_2-k_3-k_4| 
 \nonumber \\ & &-|k_1-k_2+k_3-k_4| -|k_1-k_2-k_3+k_4|  \Big). \nonumber \\ & & 
\eea
We denote by ``quad" a configuration with fixed $k_1,k_2,k_3,k_4$ that contributes to Eq.~(\ref{Testim}).
The variance of $\widetilde{T}$ is simply, 
\beq
\Delta \widetilde{T}^2= V_f \frac{s_T}{V_{\widetilde{T}}}\; P_{tot}(k_1) P_{tot}(k_2) P_{tot}(k_3) P_{tot}(k_4),
\eeq
where $s_T=24,6,2,1$ if $4,3,2$ or none of the $k_i$'s are equal or $s_T=4$ if we have two paired couples.

\section{Determination of the bias parameters}
\label{DetB}

\subsection{Simple Estimates}
\label{SEst}

We would like here to obtain a simple estimate of the uncertainty on the bias parameters that we can expect from an analysis involving bispectrum and trispectrum measurements. This is not an easy task, since the signal to noise for the angle-averaged trispectrum is given by a complicated integral that can be hardly replaced by a single, even approximate, number. What we can estimate there is the dependence on the survey volume and on the smallest scale included in the analysis, expressed in terms of $k_{\rm max}$, the maximum value of the wave number included in the sums over the configurations. \\
We consider here, as a simple example, the signal to noise due to the contribution from gravity to the bispectrum and trispectrum, i.e. that sensitive to the linear bias $b_1$, corresponding to $B$ and $T^{(1)}$ in Eqs.~(\ref{bias_bisp}) and~(\ref{Tg}). The {\em total} signal to noise in these components is then given by
\beq\label{StoNBsTs}
\left(\frac{S}{N}\right)_B^2\equiv\sum_{\rm triangles}\frac{\left[B^{(1)}\right]^2}{\Delta B^2}, \ \ \ \ \ 
\left(\frac{S}{N}\right)_T^2\equiv\sum_{\rm quads}\frac{\left[\widetilde{T}^{(1)}\right]^2}{\Delta \widetilde{T}^2},
\eeq
giving a non-marginalized uncertainty on $b_1$
\beq\label{Deltab1}
\Delta b_1^B \sim \frac{1}{(S/N)_B}, \ \ \ \ \ 
\Delta b_1^T \sim \frac{1}{2(S/N)_T},
\eeq
the factor of 2 enhancement for the trispectrum case is due to the fact that $Q_T$ is sensitive to $b_1^2$, see Eq.~(\ref{Qttot}), where we used that $B^2/\Delta B^2 \approx Q_B^2/\Delta Q_B^2$ and the same for $Q_T$. In order to make an estimate, we can replace the triple and quadruple sums with a single sum over integers introducing a coefficient that takes into account the number of configurations having the maximum of the three or four sides equal to a certain $k$ and replacing $B(k_1,k_2,k_3)$ and $\widetilde{T}^{(1)}(k_1,k_2,k_3,k_4)$ with $B(k)$ and $\widetilde{T}^{(1)}(k)$  as given by representative configurations of side $k$. Assuming a  bin in Fourier space $\delta k$, we have the {\em cumulative} signal to noise, 
\beq
\left(\frac{S}{N}\right)_B^2\simeq\sum_{i=1}^{i_{\smax}}N_B(i)\frac{\left[B(k)\right]^2}{\Delta B^2(k)},
\eeq
where $k=i\; \delta k$, $i_{\smax}=k_{\smax}/ \delta k$, $N_B(i)=i(i+1)/2$ and
\beq
\left(\frac{S}{N}\right)_T^2\simeq\sum_{i=1}^{i_{\smax}}N_T(i)\frac{\left[\widetilde{T}^{(1)}(k)\right]^2}{\Delta \widetilde{T}^2(k)},
\eeq
where $N_T(i)=i(i+1)(i+2)/6$. Now, we use
\beq
B^{(1)}(k)\simeq 3 \,Q_B\, P^2(k), \qquad \Delta B^2(k)=\frac{k_f^3\, P^3(k)}{8\pi^2 k^3 (\delta k)^3},
\eeq
with $Q_B$ some representative amplitude of order one, and for the averaged trispectrum
\beq
\widetilde{T}^{(1)}(k)\simeq C_T P^3(k), \qquad 
\Delta \widetilde{T}^2(k)=\frac{k_f^3\, P^4(k)}{32 \pi^3k^5 (\delta k)^4},
\label{CT}
\eeq
where the constant $C_T$ is difficult to determine in a simple way, being the result of complicated angular integrations, it depends on the configuration of wavevectors and on the binning size $\delta k$. 

It is instructive to compare these estimates to the power spectrum case,
\beq
\left(\frac{S}{N}\right)_P^2\simeq\sum_{i=1}^{i_{\smax}}\frac{\left[P(k)\right]^2}{\Delta P^2(k)} \sim 2 \, i_{\smax}^3 \Big( \frac{\delta k}{k_f} \Big)^3= \frac{2\, V}{(2\pi)^3}\, k_{\smax}^3,
\label{StoNPk}
\eeq
where we used $\Delta P^2(k) = 2 P^2(k)/N_i$ with $N_i=4\pi i^2 (\delta k/k_f)^3$ the number of Fourier modes in bin given by $i$, and $V$ is the volume of the survey. For the bispectrum one gets instead,
\beq
\left(\frac{S}{N}\right)_B^2\simeq 9\, \pi\, Q_B^2\ \Big( \frac{\delta k}{k_f} \Big)^3
\sum_{i=1}^{i_{\smax}}i^2 \Delta(k_i),
\eeq
where $\Delta(k)=4\pi k^3 P(k)$, assuming $Q_B \sim 1$ and an effective spectral index $n_{\rm eff} \sim -1.5$ this gives, 
\beq
\left(\frac{S}{N}\right)_B^2 \sim 6 \, i_{\smax}^3 \Big( \frac{\delta k}{k_f} \Big)^3 \Delta(k_{\smax}) =  \frac{6\, V}{(2\pi)^3}\, k_{\smax}^3 \, \Delta(k_{\smax}),
\label{StoNBk}
\eeq
Comparing Eqs.~(\ref{StoNPk}) and~(\ref{StoNBk}) we see that up to scales where $\Delta(k_{\smax}) \la 1$ {\em there should be more signal to noise in the bispectrum than the power spectrum}. This simple estimate ignores shot noise and survey geometry, but we shall see that this conclusion remains true when these are included. 

\begin{figure}
\begin{center}
\includegraphics[width=0.49\textwidth]{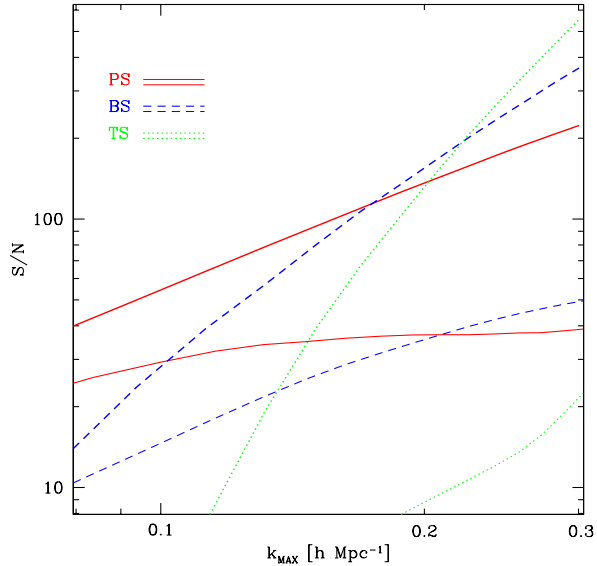}
\caption{The top three lines show the expected {\em cumulative} (up to scale $k_{\smax}$) signal-to-noise for power spectrum, bispectrum and trispectrum from an ideal survey with volume $V=0.3$ $(\Gpc)^3$ and a galaxy density $\bar{n}=0.003\; (\Mpc)^{-3}$. The bottom three lines show the same quantities for the case of the SDSS geometry (see section~\protect\ref{PsBsTs}), including radial selection function, redshift distortions, and the covariance matrix between different band powers, triangles and quads. Note that there are additional contributions due to nonlinear bias in the bispectrum and trispectrum case not included here.}
\label{StoNfig}
\end{center}
\end{figure}

To estimate the trispectrum signal to noise we would need to evaluate the constant $C_T$ in Eq.~(\ref{CT}), we will do this implicitly in Fig.~\ref{StoNfig} below when we take properly into account the sum over all configurations of the estimator in Eq.~(\ref{Testim}). For our purpose here it is enough to note that we can approximately reproduce the scaling in Fig.~\ref{StoNfig}  by using that $C_T \sim \tilde{C}_T (k/\delta k)$ with $\tilde{C}_T$ approximately constant. Then we have,
\bea
\left(\frac{S}{N}\right)_T^2 &\simeq &\tilde{C}_T^2\ \Big( \frac{\delta k}{k_f} \Big)^3
 \sum_{i=1}^{i_{\smax}} i^4 \Delta(k_i)^2 \nonumber \\ 
 &\sim & \frac{\tilde{C}_T^2\, V}{8 (2\pi)^3}   \, k_{\smax}^3 \, \Big(\frac{k_{\smax}}{\delta k}\Big)^2 \Delta(k_{\smax})^2,
\label{StoNTk}
\eea

We see that compared to the bispectrum, Eq.~(\ref{StoNTk}), the trispectrum is  suppressed  at the largest scales by a factor of $\Delta(k_{\smax})$, similar to what happens by going from the power spectrum to the bispectrum, Eq.~(\ref{StoNBk})~\cite{FMS93,Scoccimarro:1997st}.

In addition there is a bin-size dependent factor due to the particular average we are doing in our trispectrum estimator, which results from the effect of the angular integration over the PT kernels. This bin-size dependent factor is only illustrative, as we mentioned above, since it is hard to estimate precisely, but it is saying that by averaging the configuration dependence due to the PT kernels one is decreasing the signal to noise, we use below $\delta k = 3 k_f$ and therefore potentially we could gain a factor of three in signal to noise by doing ``minimal" averaging, $\delta k = k_f$. 

In order to improve over Eq.~(\ref{StoNTk}), we show in the top three lines in Fig.~\ref{StoNfig} the result of computing the signal to noise for the power spectrum (solid), bispectrum (dashed) and trispectrum (dotted) by explictly doing the respective sums over configurations up to some maximum scale $k_{\smax}$, assuming an ideal (diagonal covariance matrix) survey with volume $V=0.3$ $\Gpc^3$ and a galaxy density $\bar{n}=0.003\, (\Mpc)^{-3}$.

We see from the top three lines in Fig.~\ref{StoNfig} that the signal to noise increases faster as a function of $k_{\smax}$ for higher-order statistics, with the signal weighted more toward smaller scales. The effect of the shot noise is simple to estimate as well, for the $N-$point spectrum each scale is penalized by a factor of $[\bar{n}P_i/(1+\bar{n}P_i)]^N$, so higher-order statistics get more penalized by poor sampling. However, in the case of Fig.~\ref{StoNfig} shot noise is small enough to be almost unimportant. Given the estimates in Fig.~\ref{StoNfig},  the expected uncertainties in the linear bias from Eq.~(\ref{Deltab1}) are $\Delta b_1^B\approx 3\times 10^{-3}$ and $\Delta b_1^T\approx 10^{-3}$ for this ideal geometry. 

We should note that the signal to noise figure of merit shown in Fig.~\ref{StoNfig} does not capture the full extent of the statistical power in the bispectrum and trispectrum since there are many additional components in the presence of nonlinear bias, in fact this is the crucial point that leads to constraints on the mean of the HOD, see section~\ref{biTOhod}. The signal to noise contained in these additional terms depends on the type of galaxy, but note that nonlinear large-scale bias is inevitable in the framework of the halo model~\cite{Scoccimarro:2000gm}, even if $b_1\simeq1$ it is very difficult to have $b_2=b_3= 0$, see Eq.~(\ref{galaxy-halo-bias}) and Fig.~\ref{halobias} below.


\subsection{Likelihood Analysis: Ideal Geometry}
\label{LikeIdeal}

Since most of the signal is coming from scales small compared to the size of the survey, it is reasonable to assume that the joint likelihood for the power spectrum, bispectrum and trispectrum will be Gaussian;  this can be checked for a particular survey geometry by simulating a large pool of mock catalogs~\cite{SCJC00,Sco00b,GaSc04}. We work with the reduced amplitudes $Q_B$ and $Q_T$ which to very good approximation are independent of cosmological parameters (e.g. $\Omega_m$ and $\Omega_\Lambda$) and the amplitude of the power spectrum ($\sigma_8$). The only remaining dependence on cosmology is through the shape of the power spectrum, which we take to be fixed by power spectrum measurements here. This is only important for a relatively small number of configurations, when all scales are of the same order the reduced amplitudes $Q_B$ and $Q_T$ becomes independent of the shape of the power spectrum.  We will consider the more general case, where one simultaneously constraints the power spectrum, bispectrum and trispectrum, in future work. The joint Gaussian likelihood for $Q_B$ and $Q_T$ reads, 
\bea\label{likelihood}
-2\ln {\cal L}& =& {\rm const}+  \sum_{\rm triangles} \frac{\left(Q_B^{\rm obs}-Q_B^{\rm mod}\right)^2}{(\Delta Q_B^{\rm mod})^2}
\nonumber\\
&+& \sum_{\rm quads} \frac{\left(Q_T^{\rm obs}-Q_T^{\rm mod}\right)^2}{(\Delta Q_T^{\rm mod})^2},
\eea


where $Q_B^{\rm mod}$ and $Q_T^{\rm mod}$ are given in terms of bias parameters by Eqs.~(\ref{Qgtot}) and~(\ref{Qttot}), respectively, and by sum over ``quads" we mean those average trispectrum configurations that appear in our estimator, Eq.~(\ref{Testim}). In addition, we consider only those quads with vanishing non-connected component. In particular we limit ourselves to the cases with $k_1 > k_2 > k_3 > k_4$ (with $\kvec_1+\kvec_2+\kvec_3+\kvec_4=0$). In principle we could include configurations with two or three equal wavevector lengths, but in the case of a non trivial survey geometry such configurations have a leakage from non-connected contributions due to the coupling of the Fourier modes by the survey window. In order to be conservative we restrict here as well to these ``safe'' configurations.

Given the likelihood function in Eq.~(\ref{likelihood}), we compute the expected errors on the bias parameters using the various components of bispectrum and trispectrum as determined by Eulerian perturbation theory as described in the previous section, for surveys with ideal geometry and different volumes. We consider a $\Lambda$CDM cosmology with $\Omega_m=0.27$, $\sigma_8=0.82$, corresponding to a non-linear scale of $k_{NL}\simeq 0.3$ $\kvecMpc$. 

\begin{figure}[h]
\begin{center}
\includegraphics[width=0.4\textheight]{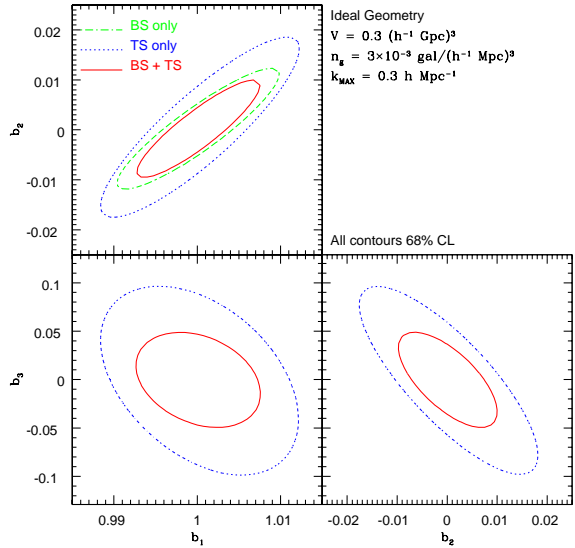}
\caption{Joint $68\%$ confidence intervals marginalized over a single parameter for $V=0.3\;(\Gpc)^3$, galaxy density $\bar{n}=3\times 10^{-3}\;(\Mpc)^{-3}$ and $k_{\smax}=0.3\;\kvecMpc$.}
\label{L0669contoursall}
\end{center}
\end{figure}

The results for three different volumes, $V=0.1$, $0.3$ and $1$ $(\Gpc)^3$, are given in table \ref{likeresultsall}. Including the trispectrum helps reduce the error on the determination of $b_1$ and $b_2$ roughly by about 20\% when all scales up to $k=0.3\;\kvecMpc$ are included. Note that one can also determine an additional bias parameter $b_3$ that would not be possible otherwise. The error on this cubic bias is not nearly as small as for $b_1$ and $b_2$, this is almost certainly related to our averaged trispectrum estimator being far from optimal for detection of $b_3$, in fact one can easily check that our average trispectrum is averaging out a significant part of the dependence on configuration shape present in the full trispectrum; improvement on this is left for future work. 
\begin{table}[t!]
\caption{\label{likeresultsall} Marginalized errors ($68\%$ CL) on the bias parameters for three survey volumes  determined using bispectrum and trispectrum alone and combined with $k_{MAX}=0.3,\;\kvecMpc$. Volumes are in $(\Gpc)^3$ and densities in $(\Mpc)^{-3}$.}
\begin{ruledtabular}
\begin{tabular}{llllll}
  V & $\bar{n}_g$ & Param. & Bisp. & Trisp. & Combined \\
\hline
    &             & $\Delta b_1$ & $0.033$  & $0.030$  & $0.030$ \\
 $1$& $10^{-4}$   & $\Delta b_2$ & $0.042$  & $0.040$  & $0.040$ \\
    &             & $\Delta b_3$ & -        & $0.18$   & $0.18$  \\
  \hline
    &             & $\Delta b_1$ & $0.0065$ & $0.0082$ & $0.0050$ \\
$0.3$&$3\!\times\!\! 10^{-3}$& $\Delta b_2$ & $0.0080$ & $0.012$  & $0.0066$ \\
    &             & $\Delta b_3$ & -        & $0.064$  & $0.032$  \\
  \hline
    &             & $\Delta b_1$ & $0.014$  & $0.025$  & $0.012$ \\
$0.11$&$10^{-3}$  & $\Delta b_2$ & $0.018$  & $0.039$  & $0.016$ \\
    &             & $\Delta b_3$ & -        & $0.21$   & $0.078$ \\
\end{tabular}
\end{ruledtabular}
\end{table}
\\
In Fig.~\ref{L0669contoursall} we give the $68\%$ confidence intervals for two bias parameters at a time marginalizing over the third, for the $V=0.3\;(\Gpc)^3$ case. 

\subsection{Likelihood Analysis: SDSS forecast}
\label{LikeSDSS}

In this section we consider a realistic survey geometry with the induced covariance matrix between different configurations both for the bispectrum and the trispectrum. We also include redshift distortions, as calculated by second-order Lagrangian Perturbation Theory (2LPT), see~\cite{Sco00b} for a comparison of 2LPT against N-body simulations for the redshift-space bispectrum, we will present a similar comparison for the trispectrum in~\cite{trispectrum}. For biasing, we assume Eq.~(\ref{Qgtot}) and ~(\ref{Qttot}) still hold in redshift space, which is a reasonable approximation near our fiducial unbiased model. 

We consider a survey geometry that approximates the north part of the SDSS survey, a $10,400$ square degree region~\footnote{See {\tt http://www.sdss.org/status} under ``spectroscopy". It corresponds to including all stripes in the north and ignoring 76-86 in the south.}.  We don't include the South part of the survey in our analysis, which has a smaller volume and a nearly two-dimensional geometry that complicates the simplified  bispectrum and trispectrum analysis we will do below.  For the radial selection function we use that from~\cite{Blanton03}, and we assume that the angular selection function is unity everywhere inside the survey region, which is a very good approximation.

The mock catalogs are the same we have used before in~\cite{SSZ04}.  Using a 2LPT code~\cite{Sco00b} with about $42\times 10^6$ particles in a rectangular box of sides $L_i=660$, $990$ and $1320 \Mpc$, we have created $6080$ realizations of the survey geometry. In all cases, cosmological parameters are as in the ideal geometry analysis and $b_1=1$, $b_2=b_3=0$. For each of these realizations, we have measured the redshift-space bispectrum and trispectrum for configurations of all shapes with sides between $k_{\min}=0.02\kvecMpc$ and $k_{\max}=0.3\kvecMpc$, giving a total of $N_{\rm triangles}=7.5\times 10^{10}$ triangles and $N_{\rm quads}=4.0\times 10^{15}$ quads. These are binned into $N_{\rm T}=1015$ triangles and $N_{\rm Q}=1720$ quads with a bin size of $\delta k=0.015\kvecMpc$. The generation of each mock catalog takes about 15 minutes, and has about $5.7\times 10^5$ galaxies. The redshift-space density field in each mock catalog is then weighed using the FKP procedure~\cite{Feldman:1993ky}, see e.g.~\cite{MVH97,Sco00b} for a discussion in the bispectrum case. The results we present correspond to a weight $P_0=5000~(\Mpc)^3$. The bispectrum and the averaged trispectrum are then measured in each realization. The bispectrum takes about $2$ minutes per realization, while the averaged trispectrum takes about 40 minutes~\footnote{Timings are for a 1.26 GHz. Pentium III processor.}. However the correction due to shot noise and geometry in the trispectrum case is nontrivial~\cite{trispectrum}, since it does not involve the power spectrum and bispectrum at the already measured configurations, but at more complicated configurations, e.g. involving $k_{12}$ instead of $k_1,k_2$. Computing these additional correlators is time consuming, though still doable, doing so by ``brute force" adds an additional 13 hours per realization.

In order to generalize the discussion given in the previous section to the case of arbitrary survey geometry, we use the bispectrum and trispectrum eigenmodes $\hat{q}_n$, see~\cite{Sco00b} for a detailed discussion for the bispectrum and~\cite{GaSc04} for the three-point function case. The discussion is the same for both, here we only summarize it for the bispectrum. The eigenmodes can be written as a linear combination (chosen here to have zero mean),

\beq
\hat{q}_n = \sum_{m=1}^{N_{\rm T}} \gamma_{mn} \frac{Q_m-\bar{Q}_m}{\Delta Q_m},
\eeq
where $\bar{Q}_m \equiv \langle Q_m \rangle$, $(\Delta Q_m)^2 \equiv \langle (Q_m-\bar{Q}_m)^2 \rangle$. By definition they  diagonalize the bispectrum covariance matrix,
\beq
\langle \hat{q}_n\, \hat{q}_m \rangle = \lambda^2_n \, \delta_{nm},
\eeq
and have signal to noise,

\beq
\left(\frac{S}{N}\right)_n = \frac{1}{\lambda_n}  \left| \sum_{m=1}^{N_{\rm T}} \gamma_{mn} \frac{\bar{Q}_m}{\Delta Q_m}\right| .
\eeq

The eigenmodes are easy to interpret when ordered  in terms of their signal to noise~\cite{Sco00b}.  The best eigenmode (highest signal to noise), say $n=1$, corresponds to all weights $\gamma_{m1}>0$; that is, it represents the overall amplitude of the bispectrum averaged over {\em all} triangles.  The next eigenmode, $n=2$, has $\gamma_{m2}>0 $ for nearly collinear triangles and $\gamma_{m2}<0$ for nearly equilateral triangles, thus it represents the dependence of the bispectrum on triangle shape. 

The same arguments hold for trispectrum eigenmodes. Here again the first eigenmode corresponds to the overall amplitude of $Q_{\widetilde{T}}$ averaged over all configurations while higher-order eigenmodes contain further information. Altough the average over the angles defining $\widetilde{T}$ in  Eq.~(\ref{Testim}) washes away a large part of the information contained in the trispectrum, we can still expect a different behavior from configurations with almost equal values for the $k_i$'s and configurations with, for instance $k_1\gg k_2,k_3,k_4$, where the average over the angles plays a little role. A more detailed analysis of the dependence of the trispectrum (full and angle-averaged) is given in~\cite{trispectrum}.

\begin{figure}[t!]
\begin{center}
\includegraphics[width=0.5\textwidth]{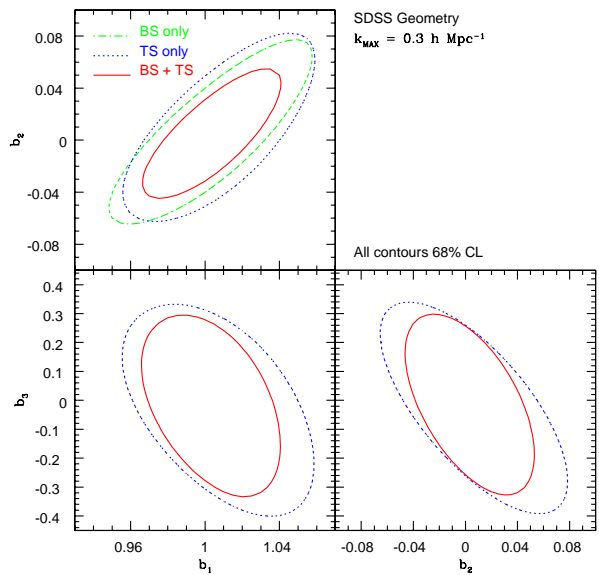}
\caption{Joint 68\% confidence intervals for two bias parameters at a time, with the third parameter marginalized over. We show results for the bispectrum (BS) only, trispectrum (TS) only, and a joint bispectrum plus trispectrum analysis. This assumes an SDSS geometry including a full covariance matrix for the bispectrum and trispectrum.}
\label{sdsspred}
\end{center}
\end{figure}

If the bispectrum and trispectrum likelihood functions are Gaussian, we can write down the likelihood for the bias parameters $b_j$ as,
\beq
{\cal L}(\{b_{j}\}) \propto \prod_{i=1}^{N_{\rm T}}\
P_{i}[\hat{q}_i^{\rm B}(\{b_{j}\})/\lambda_i^B]\times\prod_{i=1}^{N_{\rm Q}}\
P_{i}[\hat{q}_i^{\rm T}(\{b_{j}\})/\lambda_i^T],
\label{like}
\eeq
where the $P_{i}(x)$ are all equal and Gaussian with unit variance. 
We calculate the bispectrum $N_{\rm T} \times N_{\rm T} $ covariance matrix and the trispectrum $N_{\rm Q} \times N_{\rm Q} $ covariance matrix from our realizations of the survey and from that obtain the respective  $\gamma_{mn}$ and $\lambda_n$, which give the ingredients to implement Eq.~(\ref{like}). The results from such likelihood analysis are shown in table~\ref{SDSSresults} for the marginalized errors on each bias parameter and Fig.~\ref{sdsspred} for the bivariate contours with the third parameter marginalized over. The results are given for separate and joint bispectrum (BS) and trispectrum (TS). It is interesting to note that compared to the ideal geometry case in the previous section, the trispectrum helps to reduce the errors by almost $40\%$ here, twice as much as in the ideal geometry case. Again we note that the poor determination of $b_3$ compared to $b_1, b_2$ is likely due to the fact that the averaged trispectrum we use is not nearly as optimal as it could be if we use its full configuration dependence information. In addition, the trispectrum analysis by itself is expected to give similar accuracy regarding linear and quadratic bias parameters as the bispectrum. This can be used for a consistency checks of the results and sensitivity to scale dependence of the bias parameters, given that the bispectrum and trispectrum are sensitive to somewhat different scales.

\begin{table}[t!]
\caption{\label{SDSSresults} Marginalized errors ($68\%$ CL) on the bias parameters for SDSS geometry determined using bispectrum and trispectrum alone and combined with $k_{\smax}=0.3\;\kvecMpc$.}
\begin{ruledtabular}
\begin{tabular}{llll}\vspace{0.2cm}
  Parameter & Bispectrum & Trispectrum & Combined \\ 
\hline
\vspace{0.2cm}
  $\Delta b_1$ & $0.036$&$0.034$&$0.024$ \\
 \vspace{0.2cm}
  $\Delta b_2$ & $0.046$&$0.047$&$0.032$ \\
\vspace{0.2cm}
  $\Delta b_3$ & --        &$0.24$  &$0.20$  \\
\end{tabular}
\end{ruledtabular}
\end{table}

\subsection{Comparison of Signal to Noise Against the Power Spectrum: Effects of Covariance}
\label{PsBsTs}

We now go back to the question raised in section~\ref{SEst} and Fig.~\ref{StoNfig}  regarding the comparison between the signal to noise in the power spectrum, bispectrum and trispectrum. We have measured the power spectrum from the same mock catalogs and calculated the signal to noise as a function of $k_{\smax}$ from them for the power spectrum, bispectrum and trispectrum, including the effects of the covariance matrix for the SDSS geometry, always under the FKP approximation. 

The lower three lines in Fig.~\ref{StoNfig} show the results of such computation of the {\em cumulative} signal to noise~\footnote{The results in the power spectrum case agree very well with those presented in Table~3 of~\cite{Tegmark2004} when one scales their result by the ratio of survey area and takes into account that the FKP weighting is about a factor of 2 less optimal than the method used there. See e.g. discussion in~\protect\cite{HaTe00}}. We see that including the covariance matrix degrades the averaged trispectrum the most, and the power spectrum the least, as expected. The degradation in the averaged trispectrum case is rather severe, still one should keep in mind that other contributions to the trispectrum and bispectrum (due to nonlinear bias) have as much (or more) signal to noise than the gravity-only contribution displayed here.  In any case, we see that {\em higher-order statistics have comparable or larger signal to noise than the power spectrum at scales below the nonlinear scale}, as expected from the simplified analysis in section~\ref{SEst}. 

So far we have expressed the information provided by higher-order statistics in terms of constraints on bias parameters, this is the most solid (least assumptions) way of quantifying the information since it only assumes, basically, that gravity is the only long-range force in the problem. Now we discuss how these constraints can be turned into a probe of the way dark matter halos are populated with galaxies by making the additional assumption that we understand how to calculate the abundance of dark matter halos and their clustering at large scales.

\section{From Bias to HOD parameters}
\label{biTOhod}

The halo model provides a very good tool to understand galaxy biasing: in the first place the distribution of dark matter halos is related to the underlying mass distribution (\emph{halo} biasing) while the Halo Occupation Distribution (HOD) plus a radial profile prescribes how galaxies populate individual halos. While the halo distribution and halo-halo correlations can be studied and tested reliably in simulations, our understanding of galaxy clustering is still rather poor, since the non-gravitational processes involved in galaxy formation cannot be modelled accurately yet. Some of the details of how galaxies populate halos are now beginning to be explained in terms of gravitational physics (see e.g.~\cite{Zentner:2003yd} and references therein). Our purpose here is to see how much one can learn about the mean of the HOD using only large-scale information where the physics, standard gravitational instability, is better understood.

\begin{figure}
\begin{center}
\includegraphics[width=0.5\textwidth]{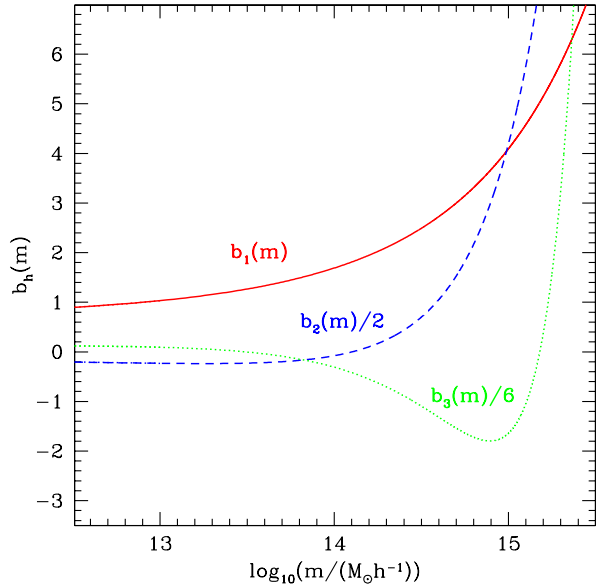}
\caption{The halo large-scale bias parameters as a function of halo mass.}
\label{halobias}
\end{center}
\end{figure}

We will assume the halo mass function to be the Sheth-Tormen (ST) mass function based on ellipsoidal collapse~\cite{Sheth:1999,Sheth:2001,Sheth:2002}, representing the average number density $n(m)$ of haloes of a given mass $m$ per unit mass. The galaxy number density is then related to the halo mass function as
\beq\label{galdens}
\bar{n}_g=\int dm\; n(m)\;\langle N_{\rm gal}(m)\rangle,
\eeq 
where $\langle N_{\rm gal}(m)\rangle$ is the mean of the HOD and it represents the average number of galaxies in an halo of mass $m$. The galaxy bias parameters are then given, in the large-scale limit, by
\beq\label{galaxy-halo-bias}
b_i= \frac{1}{\bar{n}_g}\int dm\; n(m)\; b_i(m)\;\langle N_{\rm gal}(m)\rangle,
\eeq
where $b_i(m)$ for $i=1$, $2$, $3$ are the halo large-scale bias parameters. They can be derived in the framework of non-linear perturbation theory and spherical collapse model and its extensions~\cite{Mo:1995cs,Mo:1996cn,Catelan:1997qw,Sheth:1998xe,Sheth:1999mn}; they correspond to the small $\delta$ expansion of the conditional mass function $n(m/\delta)$. In Fig.~\ref{halobias} we plot $b_1(m)$, $b_2(m)$ and $b_3(m)$ in the relevant range of masses.
The large-scale limit implies that we can ignore the size of a single halo and consider it as point-like, that is, there is no need to know the profile of galaxies inside halos. Note that in Eq.~(\ref{galaxy-halo-bias}) the bias parameters are related to halo abundance and clustering through the mean HOD, no other details about how galaxies populate halos is needed. The strategy of our approach is very simple: joint measurement of the power spectrum, bispectrum and trispectrum at large scales gives simultaneously the cosmological information to compute the halo abundance and bias, and the limits on bias parameters can be used to constraint the mean HOD through Eqs.~(\ref{galdens}-\ref{galaxy-halo-bias}).

We parametrize the mean HOD as
\beq
\langle N_{\rm gal}(m)\rangle = \left\{
\begin{array}{cc}
 0 &  {\rm for }\quad  m<M_{\rm min}\\
 1+\left(\frac{m}{M_1}\right)^\beta & {\rm for }\quad  m>M_{\rm min}
\end{array}\right.
\label{HODmean}
\eeq
where we have assumed that the average number of galaxies can be split into two contributions~\cite{Kravtsov:2003sg}: a mean occupation number for a central galaxy, corresponding to $\langle N_{cen}\rangle=1$ for $m>M_{\rm min}$ and $\langle N_{cen}\rangle=0$ for lower masses, and a mean occupation number for satellite galaxies given by $\langle N_{sat}\rangle=(m/M_1)^\beta$. We fix $M_{min}$ to be a function of $M_1$ and $\beta$ in order to reproduce the galaxy density given by Eq.~(\ref{galdens}), and then compute the joint likelihood function for $M_1$ and $\beta$ using Eq.~(\ref{galaxy-halo-bias}) and the likelihood function of the galaxy bias parameters studied in the previous section. The specific parametrization in Eq.~(\ref{HODmean }) is only chosen for illustration of our method,  one could use other parametrizations.

\begin{figure}
\begin{center}
\includegraphics[width=0.5\textwidth]{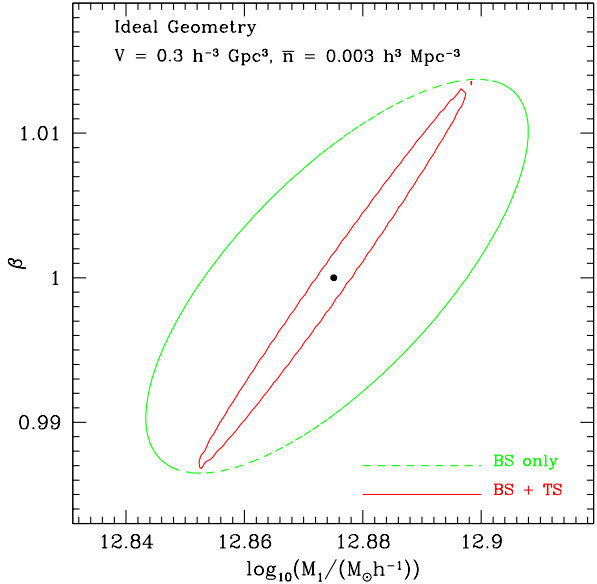}
\caption{68\% joint confidence intervals and marginalized errors for $M_1$ and $\beta$ for an ideal survey with $V=0.3(\Gpc)^3$. The larger contour corresponds to the case where only the bispectrum is used in the analysis, the inner contour includes both bispectrum and trispectrum.}
\label{contoursHODms}
\end{center}
\end{figure}

\begin{figure}
\begin{center}
\includegraphics[width=0.5\textwidth]{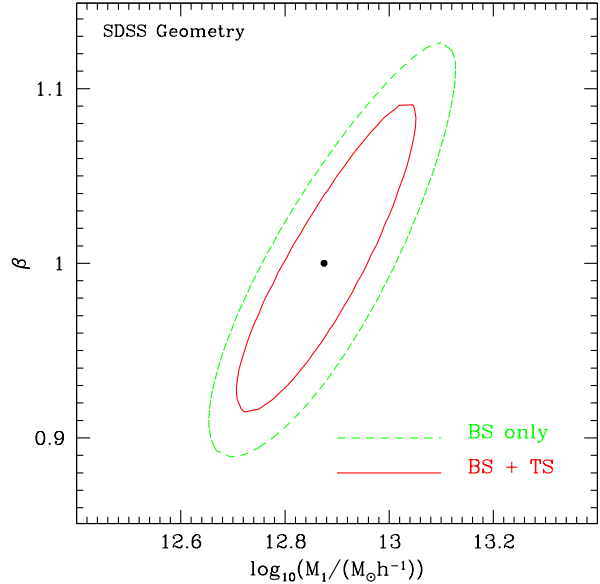}
\caption{Same as Fig.~\ref{contoursHODms} but with galaxy bias likelihood obtained from the SDSS geometry including a full bispectrum and trispectrum covariance matrix analysis.}
\label{contoursHODsdss}
\end{center}
\end{figure}

\begin{table}
\caption{\label{tabhod} Marginalized errors ($68\%$ CL) on the HOD parameters $M_1$ and $\beta$ for an ideal geometry survey with volume $V=0.3$ $(\Gpc)^3$ and a galaxy number density $\bar{n}=0.003$ $(\Mpc)^{-3}$ and for the SDSS geometry determined using bispectrum alone (in parenthesis) and bispectrum and trispectrum combined with $k_{\smax}=0.3\;\kvecMpc$.}
\begin{ruledtabular}
\begin{tabular}{ccc}
             & $\Delta \beta$      & $\Delta \log_{10}M_1$ \\
\hline 
 Ideal       & $0.0089$ ($0.0091$) & $0.016$ ($0.022$)       \\
 SDSS        & $0.059$ ($0.078$)   & $0.12$ ($0.15$)      \\
\end{tabular}
\end{ruledtabular}
\end{table}

We study the same survey examples as above, an ideal (diagonal covariance matrix) survey with volume $V=0.3$ $(\Gpc)^3$ and a galaxy number density $\bar{n}=0.003$ $(\Mpc)^{-3}$, and the  SDSS-North geometry which includes a full covariance matrix. The joint confidence intervals are presented in Figs.~\ref{contoursHODms} and~\ref{contoursHODsdss}, respectively. The fiducial values $M_1=7.5\times 10^{12} M_{\odot}/h$ and $\beta=1$ correspond to the values $b_1=0.985$ and $b_2=-0.175$ and $b_3=0.297$ for the galaxy bias parameters. In table~\ref{tabhod} we give the expected marginalized errors on $M_1$ and $\beta$. Comparing Figs.~\ref{contoursHODms} and~\ref{contoursHODsdss} one can see that for the ideal geometry the introduction of the trispectrum and thus of cubic bias information significantly  improves the determination of $M_1$, however this does not translate into a similar impact for the SDSS geometry due to the effects of the covariance matrix (expected from Fig~\ref{StoNfig}). In principle one should be able to recover a similar effect for the SDSS case by improving on the trispectrum estimator to get better constraints on $b_3$; Fig.~\ref{halobias} shows that the cubic bias of halos is a rather different function of halo mass and thus it helps to gain sensitivity on the mass scale $M_1$. 

\begin{figure}
\begin{center}
\includegraphics[width=0.49\textwidth]{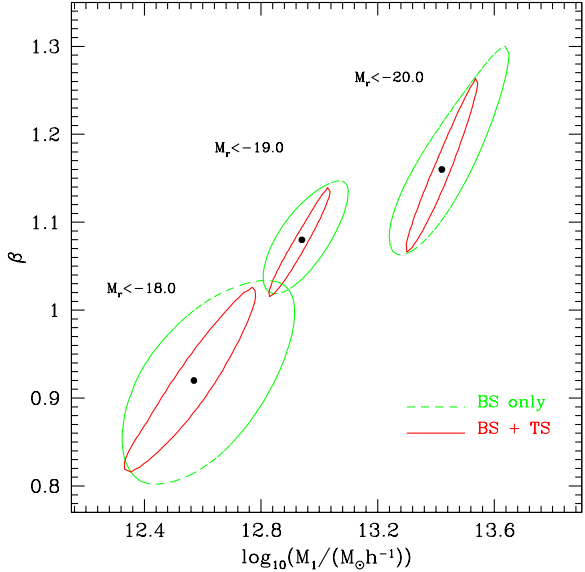}
\caption{Same as Fig.~\ref{contoursHODms} but for volumes and densities corresponding to three of the luminosity threshold samples studied in~\cite{Zehavi:2004ii}.}
\label{VLSplots}
\end{center}
\end{figure}

The example from the SDSS geometry is somewhat artificial since in a flux limited sample there is a contribution of a broad class of galaxies with different clustering properties, thus the effective bias and HOD parameters that one would obtain are not very meaningful.  Therefore we also give results (with ideal geometry) for  a series of volume limited samples studied in~\cite{Zehavi:2004ii}, see Fig.~\ref{VLSplots}. This gives an idea of how the errors on $\beta$ and $M_1$ depend on different samples and ultimately on different mass ranges. Here we assume as maximum likelihood values for $\beta$, $M_1$ and number density those given in Table~3 of~\cite{Zehavi:2004ii}, while the volumes are rescaled from $2,500$ to $10,400$ deg$^2$. Table~\ref{tabhodvls} shows the marginalized errors on $M_1$ and $\beta$ for the three subsamples.  The best constraints are expected for the sample with $M_r<-19$ since it corresponds to the best combination of volume and galaxy number density. 
  
We can compare these results with those in~\cite{Zehavi:2004ii} obtained from the two-point function analysis down to small scales. Their marginalized errors~\cite{ZhengPC} scaled to the final survey volume are better by about a factor of two for $\Delta \beta$ and three for $\log M_1$ when compared to those in Table~\ref{tabhodvls}. However, their results assume a fixed the cosmological model, and depend on further assumptions, e.g. about the galaxy profiles inside halos, modeling of the second moment of the HOD and halo-halo exclusion. Our results in Table~\ref{tabhodvls} do not include the covariance matrix, thus some degradation is expected. On the other hand, our trispectrum estimator can still be improved significantly. In the end, if further study shows that the sensitivity of our method does not compare well with the small-scale analysis approach, the interest in our method would be to provide an alternative way of probing the HOD using large-scale information that can validate the assumptions used in the small-scale analysis.

\begin{table}[t!]
\caption{\label{tabhodvls} Marginalized errors ($68\%$ CL) on the HOD parameters $M_1$ and $\beta$ for ideal geometry surveys with volumes and densities corresponding to three of the luminosity threshold samples studied in~\cite{Zehavi:2004ii}. Volumes are in $(\Gpc)^3$, densities in $(\Mpc)^{-3}$. In parenthesis we give the results from the bispectrum analysis alone.}
\begin{ruledtabular}
\begin{tabular}{ccccc}
 $M_r^{max}$ & $V$      & $\bar{n}$ & $\Delta \beta$    & $\Delta \log_{10}M_1$ \\
\hline 
 $-20$       & $0.0065$ & $ 0.006$  & $0.066$ ($0.077$) & $0.08$ ($0.14$)       \\
 $-19$       & $0.0064$ & $ 0.015$  & $0.042$ ($0.044$) & $0.08$ ($0.10$)       \\
 $-18$       & $0.0013$ & $ 0.027$  & $0.069$ ($0.076$) & $0.15$ ($0.20$)       \\
\end{tabular}
\end{ruledtabular}
\end{table}

\section{Conclusions}

We showed that current surveys have at least as much signal to noise in higher-order statistics as in the power spectrum at weakly nonlinear scales, and studied the constraints on linear, quadratic and cubic galaxy bias from measurements of the bispectrum and trispectrum at large scales. We introduced an averaged trispectrum which is relatively fast to compute in current galaxy surveys. We calculated the expected marginalized errors on the bias parameters for surveys with ideal geometry of relevant sizes and galaxy densities as well as for the more realistic geometry of Sloan Digital Sky Survey. We have shown that the trispectrum analysis alone can give at least as good results as the bispectrum in the determination on the linear and quadratic bias, which can be used for consistency checks, while in addition providing constraints on the cubic bias parameter. The combined likelihood analysis of the bispectrum and trispectrum can improve the results of the bispectrum alone by about $30$\%. 

We also discussed how one can use the bispectrum and trispectrum information to determine the mean of the galaxy halo occupation distribution (HOD), subject only to adequate modeling of the abundance and large-scale clustering of halos and thus is independent of details of how galaxies are distributed within halos. This provides a novel way of measuring the way galaxies populate halos and gives a consistency check on the traditional approach of using two-point statistics down to small scales, which necessarily makes more assumptions.

Although our results are promising, a number of checks and improvements are required to understand better the statistical power of these techniques. At the basic level, more work is needed to come up with a trispectrum estimator that is more sensitive to bias parameters and fast to evaluate, our attempt here was purely based on simplicity. Another issue is the validity of trispectrum results based on perturbation theory; comparison against $N$-body simulations in real and redshift space will be addressed elsewhere~\cite{trispectrum}. In addition, our mapping from bias parameters to constraints of the HOD assume that the halo bias parameters are well described by the Sheth-Tormen conditional mass function, but there is currently no test of these predictions for $b_2$ and $b_3$ against numerical simulations. We hope to report on this in the near future.

\acknowledgments 

We thank Andreas Berlind and Enrique Gazta\~naga for useful discussions, and Zheng Zheng  for the HOD marginalized errors from~\cite{Zehavi:2004ii}. R.~S. thanks the Kavli Institute for Cosmological Physics at the University of Chicago for hospitality during a sabbatical visit. Our work is supported by grants NSF PHY-0101738 and NASA  NAG5-12100. Our mock catalogs were created using the NYU Beowulf cluster supported by NSF grant PHY-0116590.


\begin{thebibliography}{10}

\bibitem{Percival2004}
W.~Percival et al., \mnras, {\bf 353}, 1201 (2004)

\bibitem{Tegmark2004}
M.~Tegmark et al., \apj, {\bf 606}, 702 (2004)

\bibitem{Frieman:1993nc}
J.~A.~Frieman and E.~Gazta\~naga,
Astrophys.\ J.\  {\bf 425}, 392 (1994)

\bibitem{Fry:1992vr}
J.~N.~Fry and E.~Gazta\~naga,
Astrophys.\ J.\  {\bf 413}, 447 (1993)

\bibitem{Frieman:1999qj}
J.~A.~Frieman and E.~Gaztanaga,
Astrophys.\ J.\  {\bf 521}, L83 (1999)

\bibitem{Szapudi02}
I. Szapudi et al., \apj, {\bf 570}, 75 (2002)

\bibitem{Croton04}
D.J. Croton et al., \mnras, {\bf 352}, 1232 (2004)

\bibitem{Gazta02}
E. Gazta\~naga, \apj, {\bf 580}, 144 (2002)

\bibitem{JiBo04}
Y.P. Jing, G. B\"orner, \apj, {\bf 607}, 140 (2004)

\bibitem{Kayo04}
I. Kayo et al., Pub. Astron. Soc. J., {\bf 56}, 413 (2004)

\bibitem{Fry1994}
J.~N.~Fry,
Phys.\ Rev.\ Letters {\bf 73}, 2 (1994)

\bibitem{Scoccimarro:2000sp}
R.~Scoccimarro, H.~A.~Feldman, J.~N.~Fry and J.~A.~Frieman,
Astrophys.\ J.\  {\bf 546}, 652 (2001)

\bibitem{Feldman:2000vk}
H.~A.~Feldman, J.~A.~Frieman, J.~N.~Fry and R.~Scoccimarro,
Phys.\ Rev.\ Lett.\  {\bf 86}, 1434 (2001)

\bibitem{Verde:2002ed}
L.~Verde et al., \mnras, {\bf 335}, 432 (2002)

\bibitem{FrPe78}
J.N. Fry, P.J.E. Peebles, \apj, {\bf 221}, 19 (1978)

\bibitem{MSS92}
A. Meiksin, I. Szapudi, A.S. Szalay, \apj, {\bf 394}, 87 (1990)

\bibitem{SSB92}
I. Szapudi, A.S. Szalay, P. Boschan, \apj, {\bf 390}, 350 (1992)

\bibitem{SDES95}
I. Szapudi, G.B. Dalton, G. Efstathiou, A.S. Szalay, \apj, {\bf 444}, 520 (1995)

\bibitem{BaFr91}
D.J. Baumgart, J.N. Fry, \apj, {\bf 375}, 25 (1991)

\bibitem{VeHe01}
L. Verde, A.F. Heavens, \apj, {\bf 553}, 14 (2001)

\bibitem{CSS98}
S. Colombi, I. Szapudi, A.S. Szalay, \mnras, {\bf 296}, 253 (1998)

\bibitem{SCB99}
I. Szapudi, S. Colombi, F. Bernardeau, \mnras, {\bf 310}, 428 (1999)

\bibitem{Ma:2000ik}
C.~P.~Ma and J.~N.~Fry, \mnras, {\bf 543}, 503 (2000)

\bibitem{Seljak:2000gq}
U.~Seljak,
Mon.\ Not.\ Roy.\ Astron.\ Soc.\  {\bf 318}, 203 (2000)

\bibitem{Scoccimarro:2000gm}
R.~Scoccimarro, R.~K.~Sheth, L.~Hui and B.~Jain,
Astrophys.\ J.\  {\bf 546}, 20 (2001)

\bibitem{Benson01}
A.J. Benson, \mnras, {\bf 325}, 1039 (2001)

\bibitem{White01}
M. White, L.~Hernquist, V.~Springel, \apj, {\bf 550}, L129 (2001)

\bibitem{BeWe02}
A.A. Berlind, D.H. Weinberg, \apj, {\bf 575}, 587 (2002)

\bibitem{Zheng02}
Z. Zheng, J.L. Tinker, D.H. Weinberg, A.A. Berlind, \apj, {\bf 575}, 617 (2002)

\bibitem{Berlindetal}
A.A. Berlind et al., \apj, {\bf 593}, 1 (2003)

\bibitem{Cooray:2002di}
A.~Cooray and R.~Sheth,
Phys.\ Rept.\  {\bf 372}, 1 (2002)

\bibitem{TaJa03}
M. Takada, B. Jain, \mnras, {\bf 340}, 580 (2003)

\bibitem{Kravtsov:2003sg}
A.~V.~Kravtsov, A.~A.~Berlind, R.~H.~Wechsler, A.~A.~Klypin, S.~Gottloeber, B.~Allgood and J.~R.~Primack,
Astrophys.\ J.\  {\bf 609}, 35 (2004)

\bibitem{Zentner:2003yd}
A.~R.~Zentner, A.~A.~Berlind,  J.~S.~Bullock, A.~V.~Kravtsov, R.~H.~Wechsler, 
arXiv:astro-ph/0411586

\bibitem{Scranton03}
R. Scranton, \mnras, {\bf 339}, 410 (2003)

\bibitem{Zehavi:2004ii}
I.~Zehavi et al., arXiv:astro-ph/0408569.

\bibitem{Zheng:2004id}
Z.~Zheng {\it et al.}, arXiv:astro-ph/0408564.

\bibitem{vandenBosch:2004mh}
X. Yang, H.J. Mo, Y.P. Jing, F.C. van den Bosch, Y. Chu, \mnras, {\bf 350},  1153 (2004)
 
\bibitem{Kev04}
K. Abazajian et al., arXiv:astro-ph/0408003

\bibitem{PTHalos}
R.~Scoccimarro, R.K.~Sheth, \mnras, {\bf 329}, 629 (2002)

\bibitem{Wangetal04}
Y. Wang,  X.~Yang,  H.J. Mo,  F.C. van den Bosch,  Y. Chu, \mnras, {\bf 353}, 287 (2004)

\bibitem{PTreview}
F. Bernardeau, S. Colombi, E. Gazta\~naga, R.~Scoccimarro, \physrep, {\bf 367}, 1 (2002)

\bibitem[Mandelbrot(1975)]{1975CRASM.280.1551M} 
B. Mandelbrot, Academie des Sciences Paris Comptes Rendus Serie Sciences Mathematiques, 
{\bf 280A}, 1551 (1975)

\bibitem[Peebles(1980)]{1980lssu.book.....P} 
P.~J.~E. Peebles, The Large-Scale Structure of the Universe, Princeton University Press, (1980)  

\bibitem{SC96}
I. Szapudi, S. Colombi,  \apj, {\bf 470}, 131 (1996)

\bibitem{Peebles74}
P.J.E. Peebles, \aap, {\bf 32}, 197 (1974)

\bibitem{MCSilk77}
J. McClelland, J. Silk, \apj, {\bf 217}, 331 (1977)

\bibitem{FMS93}
J.N. Fry, A.L. Melott, S.F. Shandarin, \apj, {\bf 412}, 504 (1993)

\bibitem{Scoccimarro:1997st}
R.~Scoccimarro, S.~Colombi, J.~N.~Fry, J.~A.~Frieman, E.~Hivon and A.~Melott,
Astrophys.\ J.\  {\bf 496}, 586 (1998)

\bibitem{Sco00b}
R.~Scoccimarro, \apj {\bf 544}, 597 (2000).

\bibitem{GaSc04}
E.~Gazta\~naga and R.~Scoccimarro, in preparation (2004).

\bibitem{SCJC00}
I. Szapudi, S. Colombi, A. Jenkins, J. Colberg, \mnras, {\bf 313}, 725 (2000)

\bibitem{Blanton03}
M.~R.~Blanton {\it et al.}, \apj {\bf 594}, 186 (2003)

\bibitem{Feldman:1993ky}
H.~A.~Feldman, N.~Kaiser and J.~A.~Peacock,
Astrophys.\ J.\  {\bf 426}, 23 (1994)

\bibitem{MVH97}
S.~Matarrese, L.~Verde and A.F.~Heavens,
\mnras {\bf 290}, 651 (1997).

\bibitem{trispectrum}
E.~Sefusatti and R.~Scoccimarro, in preparation.

\bibitem{SSZ04}
R.~Scoccimarro, E.~Sefusatti and M.~Zaldarriaga, \prd, {\bf 69}, 103513 (2004) 

\bibitem{Sheth:1999}
R.~K.~Sheth and G.~Tormen,
\mnras, {\bf 308}, 119 (1999)

\bibitem{Sheth:2001}
R.~K.~Sheth, H.~J.~Mo and G.~Tormen, \mnras, {\bf 323}, 1 (2001)

\bibitem{Sheth:2002}
R.~K.~Sheth and G.~Tormen,
Mon.\ Not.\ Roy.\ Astron.\ Soc.\  {\bf 329}, 61 (2002)

\bibitem{Mo:1995cs}
H.~J.~Mo and S.~D.~M.~White,
Mon.\ Not.\ Roy.\ Astron.\ Soc.\  {\bf 282}, 347 (1996)

\bibitem{Mo:1996cn}
H.~J.~Mo, Y.~P.~Jing and S.~D.~M.~White, \mnras, {\bf 284},189 (1997)

\bibitem{Catelan:1997qw}
P.~Catelan, F.~Lucchin, S.~Matarrese and C.~Porciani, \mnras, {\bf 297}, 692 (1998)

\bibitem{Sheth:1998xe}
R.~K.~Sheth and G.~Lemson, \mnras, {\bf 304}, 767 (1999)

\bibitem{Sheth:1999mn}
R.~K.~Sheth and G.~Tormen,
Mon.\ Not.\ Roy.\ Astron.\ Soc.\  {\bf 308}, 119 (1999)

\bibitem{ZhengPC} 
Z. Zheng, private communication (2004)

\bibitem{HaTe00}
M.~Tegmark and A.J.S.~Hamilton, \mnras, {\bf 312}, 285 (2000)

\end{thebibliography}
\end{document}